\documentclass[a4paper,fleqn,usenatbib]{mnras}
\usepackage{newtxtext,newtxmath}

\usepackage[T1]{fontenc}
\usepackage{ae,aecompl}

\usepackage{graphicx}	
\usepackage{amsmath}	

\title[ML approach for state classification]
{A Machine Learning approach for classification of accretion states of Black hole binaries}

\author[Sreehari \& Anuj Nandi]{H. Sreehari$^{1}$\thanks{E-mail: hjsreehari@gmail.com (HS), 
anuj@ursc.gov.in (AN)}, Anuj Nandi$^{2}$\\
1. Indian Institute of Astrophysics, Bangalore, 560034, India.\\
2. Space Astronomy Group, ISITE Campus, U. R. Rao Satellite Centre, Outer
Ring Road, Marathahalli, Bangalore, 560037, India.
}

\date{Accepted XXX. Received YYY; in original form ZZZ}

\pubyear{2020}

\begin{document}
\label{firstpage}
\pagerange{\pageref{firstpage}--\pageref{lastpage}}
\maketitle

\begin{abstract}

In this paper, we employ Machine Learning algorithms on multi-mission observations
for the classification of accretion states of outbursting Black hole X-ray binaries for the first time.
Archival data from {\it RXTE}, {\it Swift}, {\it MAXI} and {\it AstroSat} observatories are used to generate the 
hardness intensity diagrams (HIDs) for outbursts of the sources XTE J1859+226 (1999 outburst), GX 339-4 (2002, 2004, 2007 and 2010 outbursts), IGR J17091-3624 (2016 outburst),  and MAXI J1535-571 (2017 outburst). Based on variation of X-ray flux, hardness ratios, presence of various types of Quasi-periodic Oscillations (QPOs), photon indices and disk temperature, we apply clustering algorithms like K-Means clustering and Hierarchical clustering to classify the accretion states (clusters) of each outburst. As multiple parameters are involved in the classification process, we show that clustering algorithms club together the observations of similar characteristics more efficiently than the `standard' method of classification. We also infer that K-Means clustering provides more reliable results than Hierarchical clustering. We demonstrate the importance of the classification based on machine learning by comparing it with results from `standard' classification. 
\end{abstract}

\begin{keywords}
methods: data analysis - accretion, accretion disc - black hole physics - X-rays: binaries.
\end{keywords}



\section{Introduction}
\label{s:intro}
Black hole X-ray binaries (BH-XRBs) are binary star systems harbouring a black hole along with a secondary (donor) star.
These systems undergo periods of high X-ray activity called outbursts extending from several days to a few months in between the quiescent states \citep[and references therein]{Remillard2006,Belloni2010,Nandi2012,Nandi2018,Sreehari2018,Baby2020}. During the outburst phase, the X-ray intensity increases gradually up to a peak value and then decays down until the source goes into quiescence. These sources are known to exhibit rich variability in temporal features \citep[and references therein]{Belloni2000,Belloni2005,Nandi2012,Belloni2014,Radhika2018,Sreehari2019a}. One of the salient features exhibited by these sources are QPOs \citep{Van1985,Casella2004,Nandi2012,Sreehari2019,Ingram2019}. QPOs are asymmetric peaks in the power density spectra of XRB sources. These features are modelled using \texttt{lorentzians} and characterised using quality factor (Q), rms and significance \citep[see][for details]{Casella2005,Sreehari2019}.
QPOs with frequency in the range $0.1 - 40$~Hz are considered as Low frequency QPOs \citep[LFQPOs,][and references therein]{Casella2004,Motta2011,Yadav2016} and those above 40 Hz are considered High frequency QPOs \citep[HFQPOs,][]{Morgan1997,Remillard1999,Strohmayer2001,Altamirano2012,Belloni2013,Belloni2019,Sreehari2020}. LFQPOs are further classified into Type-A, Type-B and Type-C  \citep{Casella2005}. Type-C QPOs have higher rms ($3 - 16$~\%) and Q ($7 - 12$), and are usually observed in the harder spectral states, whereas Type-B QPOs with $2 - 4$~\% rms and Q $\ge$ 6 are found in the intermediate spectral states \citep{Belloni2005,Motta2011,Nandi2012}.

\begin{table*}
	\caption{Summary of observations used in the present work.}
	\begin{tabular}{|l|c|c|c|}
		\hline 
		Source & Observatory/Instrument & Years & References\\
		\hline
		XTE~J1859+226   & {\it RXTE/PCA+HEXTE}  & 1999 & \cite{Casella2004,Radhika2016,Nandi2018}\\
		GX~339-4        & {\it RXTE/PCA+HEXTE}  & 2002, 2004, 2007 \& 2010 & \cite{Belloni2005,Belloni2006,Nandi2012,Aneesha2019}\\
		IGR~J17091-3624 & {\it Swift/XRT}       & 2016 & \cite{Xu2017,Radhika2018}\\
		MAXI~J1535-571  & {\it MAXI, AstroSat/LAXPC, Swift/XRT} & 2017 & \cite{Tao2018,Sreehari2019}\\
		\hline
	\end{tabular}
	\label{tab:obs}
\end{table*}

\begin{figure*}
	\begin{center}
		\includegraphics[trim=0 0 0.25mm 0, clip = true, width=0.45\textwidth, height=6.5cm]{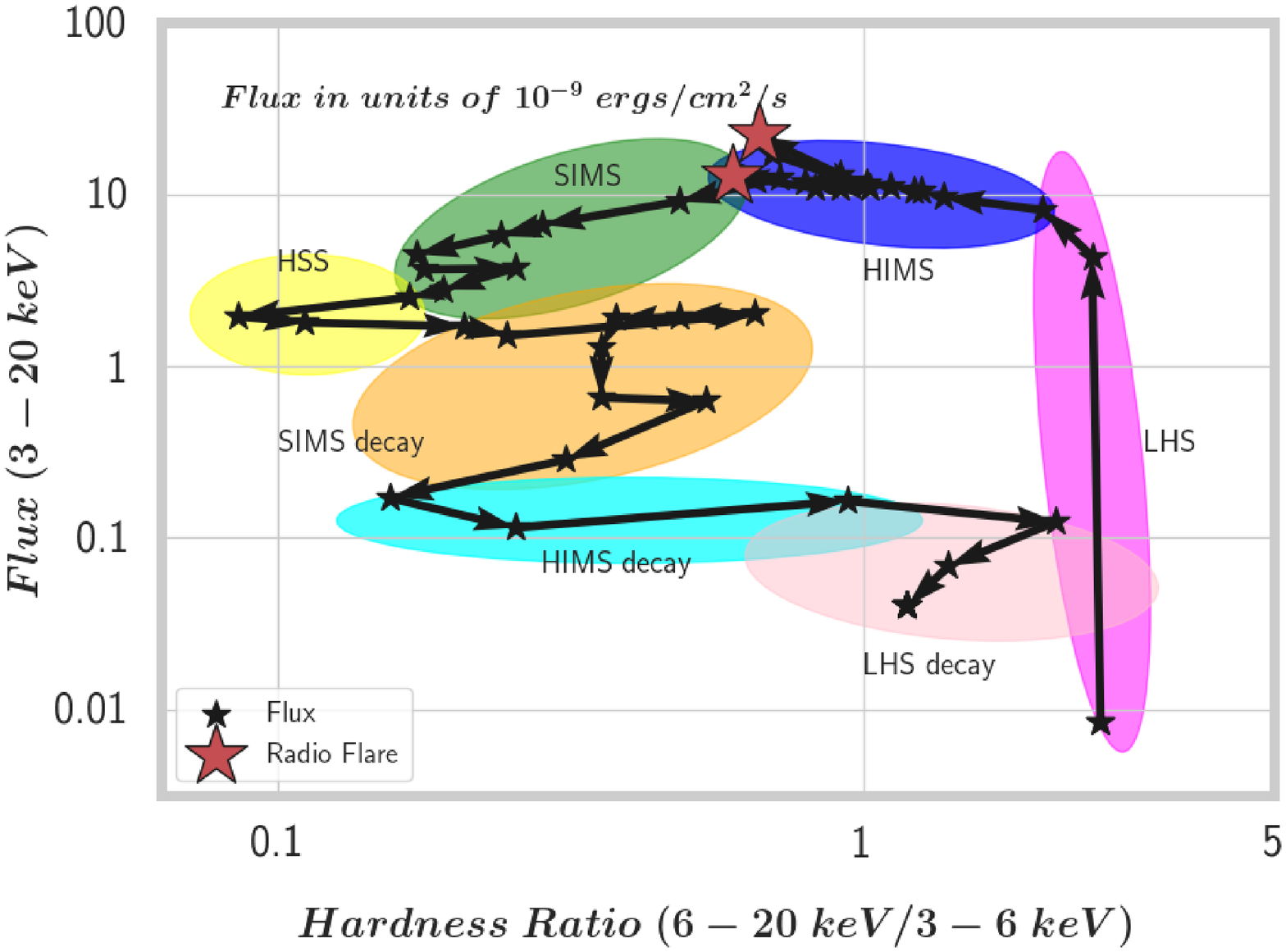}\
		\includegraphics[trim=0 0 0.25mm 0, clip = true, width=0.45\textwidth, height=6.5cm]{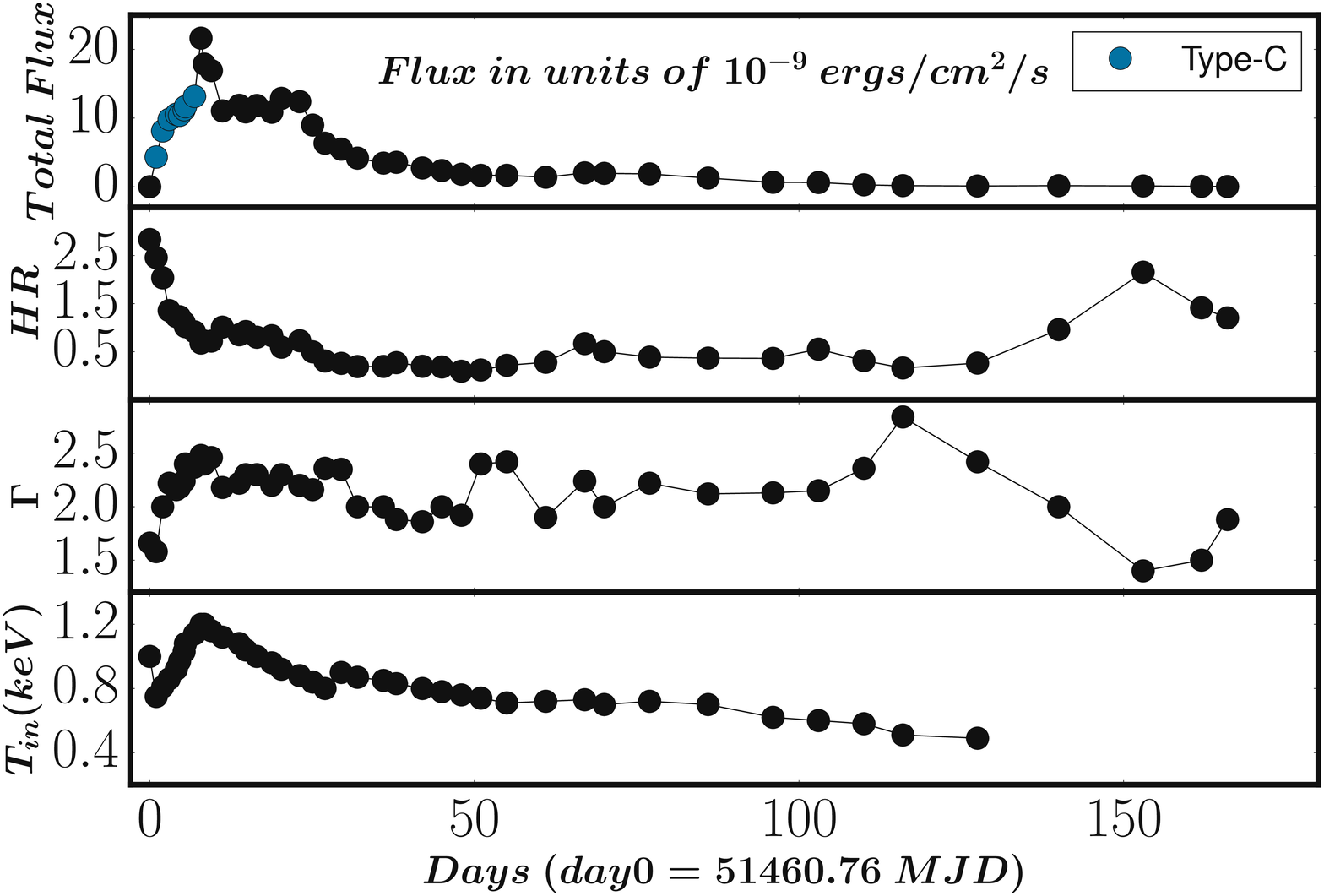}
	\end{center}
	\caption{The left plot shows the HID of the 1999 outburst of XTE J1859+226 with LHS in magenta, HIMS in blue, SIMS in green, HSS in yellow, SIMS-decay in orange, HIMS-decay in cyan and LHS-decay in pink. The radio flares are represented by red stars. The same colour convention for accretion states is followed through out this paper. The evolution of physical parameters obtained from spectral fitting are presented on the right plot. The four panels are Total flux ($3 - 20$~keV), Hardness Ratio ($6 - 20$~keV/$3 - 6$~keV), Photon Index ($\Gamma$) and the inner disk temperature (kT$_{\rm in}$). In the top panel, the presence of Type-C QPOs are indicated in blue colour.}
	\label{fig:XTEJparams}
\end{figure*} 

In addition to the temporal features mentioned above, we can infer more about the accretion states of the black hole binaries by modelling its energy spectra. 
Typically, the nature of energy spectra of BH-XRBs are characterised by a multi-temperature blackbody emission that originates from a Keplerian disc \citep{SS1973}. This accounts only for the thermal emission from the source represented by the inner disc temperature (${\rm kT}_{\rm in}$). The high energy emission from the BH-XRB is accounted for by the inverse-Comptonization \citep{Titarchuk1994,Tanaka1995,Zd1996} component represented by \texttt{powerlaw} based models  \citep{Nandi2012,Iyer2015,Radhika2016a} having a photon-index ($\Gamma$) parameter.

In general, a Hardness-Intensity Diagram (HID or q-diagram) provides a better idea about the accretion state transitions in a BH-XRB  \citep{Homan2001,Homan2005,Belloni2005,Nandi2012,Radhika2016a,Sreehari2019,Baby2020}. It depicts the variation of total flux versus the hardness ratio during an outburst (see left panel of Figure \ref{fig:XTEJparams}). The outburst duration is classified into different spectral states based on factors like hardness ratio (ratio of high energy flux to low energy flux; hereafter HR), presence or absence of QPOs, and photon-index ($\Gamma$) of the energy spectra along with the disk temperature (${\rm kT}_{\rm in}$). For instance, right panel of Figure \ref{fig:XTEJparams} shows the parameters on which the accretion state
of the 1999 outburst of the source XTE J1859+226 depends on. 
Various branches within the q-diagram corresponding to this outburst are classified into Low/Hard state (LHS), Hard-Intermediate State (HIMS), Soft-Intermediate State (SIMS) and High/Soft State (HSS) during the rising phase. The reverse trend is observed during the decay phase where the source transits to SIMS-decay, HIMS-decay and finally to LHS-decay before reaching the quiescent phase (see Figure \ref{fig:XTEJparams} and \cite{Nandi2018}). 
It is observed that the LHS corresponds to maximum HR values accompanied by a low value of flux.
The flux usually reaches a high value at the end of this state or the beginning of the HIMS.
In the HIMS, the flux value remains the same though the hardness decreases. Generally, Radio flares (i.e. transient jets) are observed at the transition region between HIMS and SIMS \citep[and references therein]{Fender2004,Fender2009,Radhika2014,Radhika2016}. As the source transits through the SIMS, the HR further decreases until the HSS is reached. In the decay phase, the flux level remains low and the HR gradually increases
as the source transits through SIMS, HIMS and LHS. Type-C QPOs are generally associated with LHS and HIMS. Their frequency values are observed to increase during the rising phase and to decrease as the source transits through the decay phase \citep{Rod2002,Chakra2008,Nandi2012,Nandi2018,Sreehari2019a}. 
Generally, the photon index ($\Gamma$) of energy spectra increases from 1.4 to $\sim$3 \citep{Remillard2006,Belloni2010,Nandi2012}
as the accretion state evolves from hard to soft through the intermediate states. 
It is observed that the inner disk temperature (${\rm kT}_{\rm in}$) is above 1~keV during the soft states \citep{Tomsick1999,Radhika2018,Baby2020}.  
In general, the energy spectra are dominated by the Comptonization component during the hard states (LHS and HIMS)
whereas in SIMS and HSS it is dominated by thermal emission \citep{SS1973,Titarchuk1994,Tanaka1995,Giannios2005,Nandi2012,Sreehari2020}.

Though the HID comes in handy to make a good classification, the fact that the accretion states depend on more physical parameters besides the flux, hardness ratio, presence and type of QPOs make the standard classification tedious and time consuming. 
For instance, it would be better to account for the inner disk temperature (kT$_{\rm in}$) and the photon index ($\Gamma$) from the energy spectral modelling while carrying out the classification. That is, if there are `N' factors like HR, flux, QPOs, kT$_{\rm in}$ and $\Gamma$ on which the classification depends, then we can have an N-dimensional space in which the data is distributed. Now, it is possible to classify each group of points (observations) having similar characteristics in this space separately by accommodating a machine learning approach.

In this paper, we attempt to make the classification of accretion states during seven outbursts (single outbursts from XTE J1859+226, IGR J17091-3624 \& MAXI J1535-571 and multiple outbursts from GX 339-4) of BH-XRBs using Machine Learning techniques for the first time. Specifically, we intend to use distance based methods like the K-means clustering \citep{Macqueen1967} and the hierarchical clustering \citep{Ward1963} algorithms for the classification of accretion states. Prior to this, Machine Learning methods have been employed in astronomy for several scenarios. For instance, the K-means clustering of SDSS (Sloan Digital Sky Survey) galaxy spectra \citep{Sanchez2010} were used to categorize galaxies with similar spectral properties together. Unsupervised clustering algorithm was used to categorise supernovae \citep{Rubin2016} light curves into slow-rising, fast-rise/slow-decline and fast-rise/fast-decline groups. \cite{Huppenkothen2017} attempted the categorisation of variability classes of GRS 1915+105 using principal component analysis (PCA) followed by logistic regression.
A rejection algorithm based on machine learning is employed to detect exo-planet transits \citep{Mislis2018} in large-scale survey data. \cite{Tei2018} used an Artificial Neural Network (ANN) method to distinguish between Active Galactic Nuclei (AGNs) and star forming galaxies. \cite{Carruba2019} used Hierarchical clustering algorithm to group asteroids into different asteroid-families based on their distance from the reference body.
Three classes of Gamma-ray bursts (GRBs) were identified using K-Means clustering algorithm by \cite{Chattopad2007}.
Recently, \cite{George2019} applied the concept of recurrence networks to classify binary stars into semi-detached, overcontact and ellipsoidal binaries.
A few other automated methods are also worth mentioning in this context. 
\cite{Dunn2010} used a decision tree based automated algorithm to model energy spectra and obtain the best fitting model.
\cite{Tetarenko2016} generated machine readable tables to create a database of Galactic BH-XRBs. A novel method for X-ray binary variability comparison called the `power colours' method was introduced by \cite{Heil2015}.	 
The present work is based on two fundamentally different clustering algorithms using which we classify the accretion states of Black hole binaries.

This paper is organised as follows.
In \S \ref{s:obs}, we detail the observations used in this paper and the standard approach for spectral state
classification of outbursting BH-XRBs. We introduce the machine learning based algorithms like K-Means clustering and Hierarchical
clustering in \S \ref{s:ML_clustering}. The results from the application of these two algorithms for multiple BH sources are presented in \S \ref{s:Results}. Finally, we conclude after discussing the results in \S \ref{s:Disc}.
  

\section{Observations and Approach}
\label{s:obs}
We re-analyse the archival data from HEASARC\footnote{\url{https://heasarc.gsfc.nasa.gov/cgi-bin/W3Browse/w3browse.pl}} and ISSDC\footnote{\url{https://astrobrowse.issdc.gov.in/astro\_archive/archive/Home.jsp}} of multiple outbursting BH-XRBs using various X-ray observatories. 
Altogether, we use data of seven outbursts (see Table \ref{tab:obs}) from four different black hole binary sources.
Four of these are from the source GX 339-4 which was active multiple times during the {\it RXTE} era\footnote{\url{https://heasarc.gsfc.nasa.gov/docs/xte/XTE.html}}. 
The other sources are XTE J1859+226 ({\it RXTE}), IGR J17091-3624 ({\it Swift/XRT}) and MAXI J1535-571 ({\it MAXI, XRT} \& {\it AstroSat}).

\subsection{Timing Analysis and Modelling}
We use archival data of {\it RXTE} for analysing the 1999 outburst of XTE J1859+226 and multiple outbursts (2002, 2004, 2007 and 2010) of GX 339-4. {\it RXTE/PCA} light curves in the energy range $3 - 20$~keV are extracted for these outbursts following \cite{Radhika2014} and \cite{Sreehari2019a}. The power spectra for these light curves are generated using the \texttt{powspec} tool of the \texttt{XRONOS}\footnote{\url{https://heasarc.gsfc.nasa.gov/xanadu/xronos/xronos.html}} package. The power spectral features (QPOs and continuum) are modelled with multiple lorentzians. QPOs are identified based on the value of quality factor (${\rm Q} = \nu/FWHM$), rms and significance ($\sigma=norm/(negative\_error)$). \texttt{Lorentzian} features with Q and $\sigma$ above 3 are generally considered as QPOs. Similarly, QPO parameters for the 2016 outburst of the source IGR J17091-3624 are obtained from \cite{Radhika2018}. For the source MAXI J1535-571, we have used QPO information from {\it AstroSat} and {\it Swift/XRT} following \cite{Sreehari2019}. We use presence of different types of QPOs along with flux and HR values as inputs to our Machine Learning algorithm in order to classify the spectral states of each source.

\subsection{Spectral Analysis and Modelling}
We follow the standard data extraction procedure \citep{Sreehari2019a} of {\it RXTE/PCA+HEXTE} to extract spectra in the range $3 - 150$~keV for GX 339-4 and XTE J1859+226. Similarly, the {\it Swift/XRT} spectra are extracted following \cite{Radhika2018}. The energy spectra thus extracted are modelled using \texttt{phabs(diskbb+smedge$\times$powerlaw)} following \cite{Radhika2014} and \cite{Aneesha2019}. The total flux in $3 - 20$~keV corresponding to the best fitting model is used to generate the HID. In the case of Swift/XRT data, we use flux in the energy range $0.5 - 10$~keV. Additionally, we use inner-disc temperature (${\rm kT}_{\rm in}$) from \texttt{diskbb} and photon index ($\Gamma$) from \texttt{powerlaw} as inputs for our analysis. For the source MAXI J1535-571, we have used the flux and hardness ratios obtained from {\it MAXI} webpage\footnote{\url{http://maxi.riken.jp/mxondem/}} following \cite{Sreehari2019}.

Following the spectral data modelling, we compute the HR as the ratio of flux in higher energy band ($6 - 20$~keV) to the flux in lower energy band ($3-6$~keV). For the case of {\it Swift/XRT}, this ratio is calculated considering $4 - 10$~keV and $0.5 - 4$~keV bands.
In the `standard' classification, the right branch of a q-diagram (see left panel of Figure \ref{fig:XTEJparams}) corresponds to LHS, the top branch corresponds to HIMS, the top left branch corresponds to SIMS and the left branch represents HSS. In the decay phase with lower flux values, the source transits through SIMS, HIMS and LHS corresponding to an increase in HR \citep{Belloni2005}. It is observed that Type-C QPOs are generally found in the LHS and the HIMS \citep{Motta2011}, whereas Type-B QPOs occur mostly in the SIMS. Radio flares are often observed in the transition region between the HIMS and SIMS \citep{Fender2009,Radhika2014}. Disc emission is significant in the softer states (SIMS and HSS) and photon index ($\Gamma$) increases as the source becomes softer. 

As we have to simultaneously take into account several parameters mentioned above in order to classify the accretion states, 
it would be better if we develop a standard algorithm to carry out this classification. In this paper, we attempt to incorporate all these factors (flux, HR, presence of QPOs, photon index, inner disk temperature) along with the observation time (MJD), in order to get a faster, easier and more reliable classification scheme based on clustering techniques.

\begin{figure}
	\begin{center}
		\includegraphics[trim=0 0 0.25mm 0, clip = true, width=0.45\textwidth]{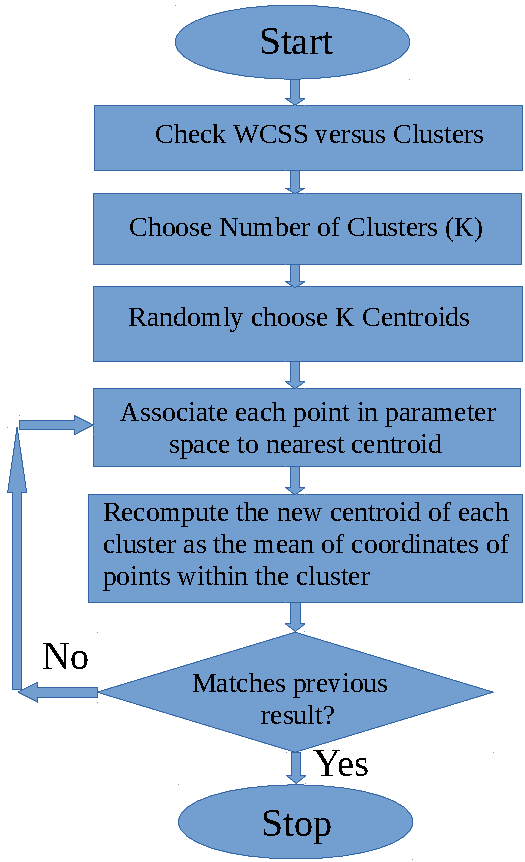}
	\end{center}
	\caption{The flowchart of K-Means clustering algorithm is shown. WCSS (Within Cluster Sum of Squared) distances is the sum of squared distances between each point in the clusters and the corresponding cluster centroid. Once the number of clusters (K) are chosen, the algorithm starts with K randomly chosen centroids. Each point is associated to the nearest centroid and then the centroid is updated as the mean of coordinates of the points within the cluster. This process is repeated until consistent results are obtained.}
	\label{fig:k-Means_flowchart}
\end{figure}

\begin{figure}
	\begin{center}
		\includegraphics[trim=0 0 0.25mm 0, clip = true, width=0.45\textwidth]{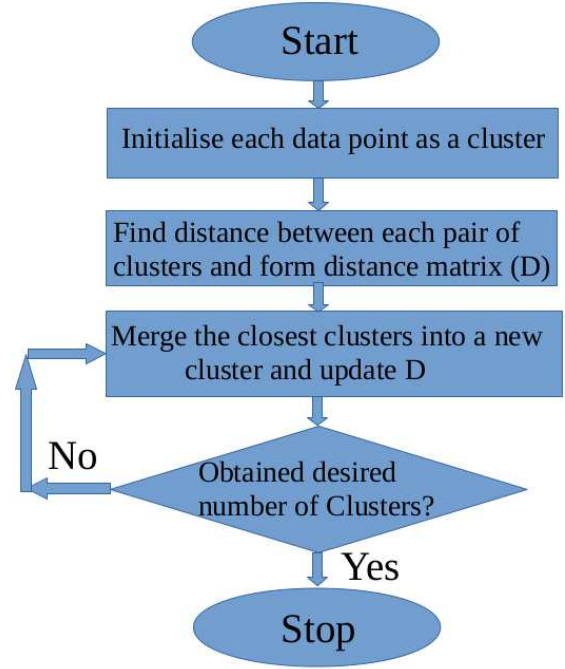}
	\end{center}
	\caption{The flowchart of Hierarchical clustering algorithm is shown. To begin with, each data point has been considered as a cluster. Then distance between each pair of points (clusters) is found and stored as a matrix (D). The points separated by the shortest distance are then merged to form a new cluster and D is updated. Again, the clusters separated by the shortest distance are identified and merged together based on the complete linkage criterion. This process is repeated until we get the desired number of clusters. See text for details.}
	\label{fig:Hier_flowchart}
\end{figure}

\section{Clustering Algorithms}
\label{s:ML_clustering}
In this section, we introduce two of the commonly used clustering techniques in machine learning that we will employ
to carry out the classification of accretion states of BH-XRBs. The algorithms that we use are K-Means clustering \citep{Macqueen1967} and Hierarchical
clustering \citep{Ward1963}. We classify accretion states for seven outbursts with both these methods and compare it with the `standard' results. Our intention is to identify the clustering method that is more suitable to address the accretion state classification problem.

In order to pre-process the data to be clustered, we have normalized all numerical parameters like
time (MJD), flux, hardness ratio (HR), photon index ($\Gamma$) and temperature (kT$_{\rm in}$). After this, we convert the categorical (non-numerical) entries like presence of QPOs and Instrument used for observations (eg: {\it RXTE/PCA}, {\it Swift/XRT}) into dummy variables using the \textit{pandas} \citep{mckinney2010} function \textit{get\_dummies}. This function performs a `one-hot' encoding on the categorical data. In `one-hot' encoding, we represent a categorical variable which can take four values (say) as a vector of length four. Suppose we have QPO's column in our data which can take values `A', `B', `C' and `No' corresponding to Type-A, Type-B, Type-C and No\_QPOs respectively. If only a Type-C QPO is present in an observation, the `one-hot' encoded vector representing QPOs for that observation is [0, 0, 1, 0], where all entries are zero except for the position indicating a Type-C QPO.  
The input parameters for the clustering algorithm used in each case are chosen based on the Explained Variance Ratio \citep[EVR,][]{Achen1982}. The EVR is the ratio of variance of a parameter to the net variance of all the parameters. The variations in the output variable are influenced mostly by the parameters which have the largest EVR. 
The same pre-processed data are used as inputs for both K-Means and Hierarchical clustering, which are described below.

\subsection{K-Means Clustering}
\begin{figure}
	\begin{center}
		\includegraphics[trim=0 0 0.25mm 0, clip = true, width=0.45\textwidth, height=6.5cm]{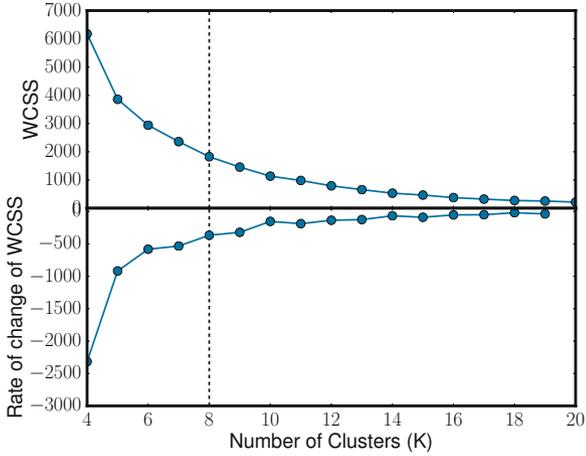}
	\end{center}
	\caption{Within cluster sum of squared (WCSS) distances versus number of clusters. The elbow of this plot is to be identified and a perpendicular is to be drawn from the elbow on to the X-axis to obtain the required number of clusters. The plot indicates the number of clusters to be around 8. The bottom panel shows the derivative of WCSS w.r.t. number of clusters. It is evident that the curve flattens around 8 clusters indicating the optimal number of clusters required. See text for details.}
	\label{fig:k-Means_WCSS}
\end{figure}

We have used the scikit-learn \citep{Pedregosa2011} implementation of K-Means clustering \citep{Macqueen1967}
algorithm to classify the spectral states of BH-XRBs. The flowchart for this algorithm is presented in Figure \ref{fig:k-Means_flowchart}. Given a set of points in an N-dimensional space,
the K-Means algorithm starts by randomly (based on a seed) guessing K (must be pre-defined) centroids.
The algorithm finds the distance to each point from each of the centroids and groups each point to
the nearest centroid. At this instance, we get K groups or clusters. Now each centroid is updated to be
the mean value coordinate within each group. Then the distances to each point from the new centroid are re-computed
and the cluster is updated. This process continues until we get same centroids and clusters in consecutive iterations. 
The value of K can be an informed guess or it can be obtained from the elbow method (see Figure \ref{fig:k-Means_WCSS}).
In the elbow method, the Within Cluster Sum of Squared (WCSS) distances for several number of
clusters is calculated and the perpendicular from the `elbow' of the plot is drawn on to the X-axis.
The elbow is the region of this plot where the rate of change of WCSS with respect to the number of clusters reduces considerably.
The point where the perpendicular from the elbow meets the X-axis is chosen as K as shown in Figure \ref{fig:k-Means_WCSS}.

\subsection{Hierarchical Clustering}
Unlike K-Means clustering, Hierarchical clustering \citep{Ward1963,Duda2000} is based on a bottom-up algorithm.
The algorithm is initiated with each point representing a cluster. At every further iteration the algorithm merges
the nearest clusters together. The distance calculation between clusters can be based on different linkages namely
single, complete, ward and average. We use complete linkage where the distance between two clusters is the largest point-wise separation in between clusters. The process stops once we get the desired number of clusters. This method is computationally more expensive because of the requirement of a linkage function besides the distance measure. Flowchart for this method is presented in Figure \ref{fig:Hier_flowchart}. 

In the following section, we present the results of clustering methods when applied to the scenario of classification of accretion states in BH-XRB sources.

\section{Clustering Outcomes}
\label{s:Results}

In this section, we present the outcomes of the application of clustering algorithms to accretion state classification of
black hole binaries. Unless otherwise mentioned, we present only one result based on which algorithm produced the better classification corresponding to each outburst. 

\begin{table*}
	\caption{The Explained Variance Ratios (EVRs) of parameters for each outburst considered in this paper are presented here. For the source MAXI J1535-571, we have used {\it MAXI} flux to generate the complete q-diagram and hence spectral parameters were not taken into account.}  
	\begin{tabular}{|c|c|c|c|c|c|c|c|}
		\hline
		Source & Outburst & MJD & Flux & HR & $\Gamma$ & ${\rm kT}_{\rm in}$&Type-C/Type-B\\
		\hline
		XTE J1859+226 & 1999 & 0.139 & 0.096 & 0.099 & 0.060 & 0.106&0.248\\
		GX 339-4      & 2002 & 0.158 & 0.107 & 0.128 & 0.067 &   -  &0.268\\
		GX 339-4      & 2004 & 0.175 & 0.083 & 0.200 & 0.109 & 0.139&0.145\\
		GX 339-4      & 2007 & 0.099 & 0.119 & 0.170 & 0.130 &    - &0.240\\
		GX 339-4      & 2010 & 0.170 & 0.087 & 0.169 & 0.045 & 0.116&0.204\\
		IGR J17091-3624&2016 & 0.090 & 0.059 & 0.097 & 0.121 & 0.196&0.133\\
		MAXI J1535-571 &2017 & 0.195 & 0.120 & 0.051 &  -    &   -  &0.116/0.211\\
		\hline
	\end{tabular}
	\label{tab:EVR}
\end{table*}

\subsection{XTE J1859+226: 1999 Outburst}


\begin{figure}
	\begin{center}
		\includegraphics[trim=0 0 0.25mm 0, clip = true, width=0.45\textwidth, height=6.5cm]{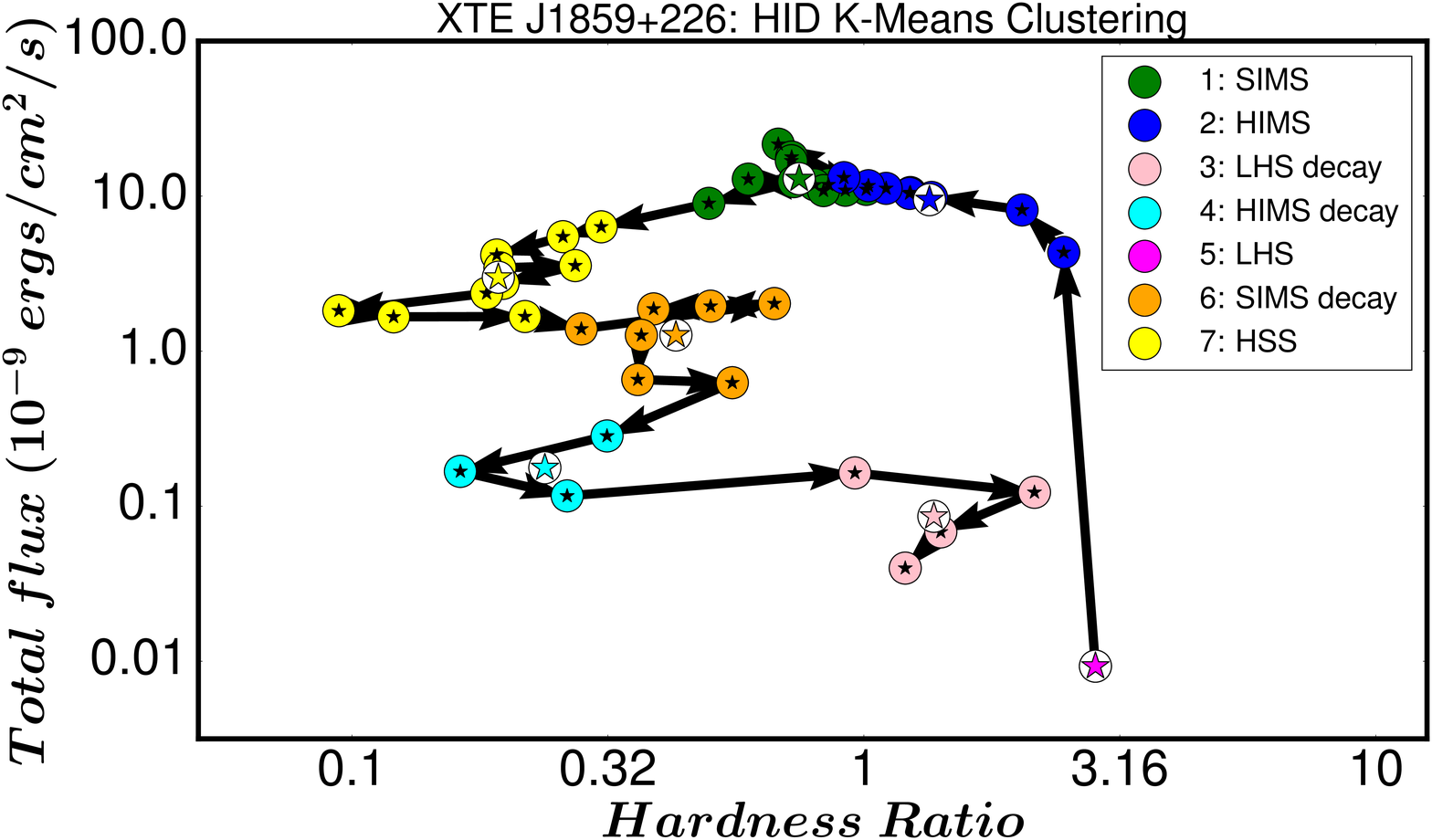}
	\end{center}
	\caption{Outcome of K-Means clustering of the q-diagram for the 1999 outburst of XTE J1859+226. Coloured stars within white circles indicate the position of centroid of each cluster. The legend is provided in the order in which each cluster is identified. The colour convention is the same as that used in Figure \ref{fig:XTEJparams}.}
	\label{fig:k-Means_XTE}
\end{figure}


\begin{figure}
	\begin{center}
		\includegraphics[trim=0 0 0.25mm 0, clip = true, width=0.5\textwidth, height=7cm]{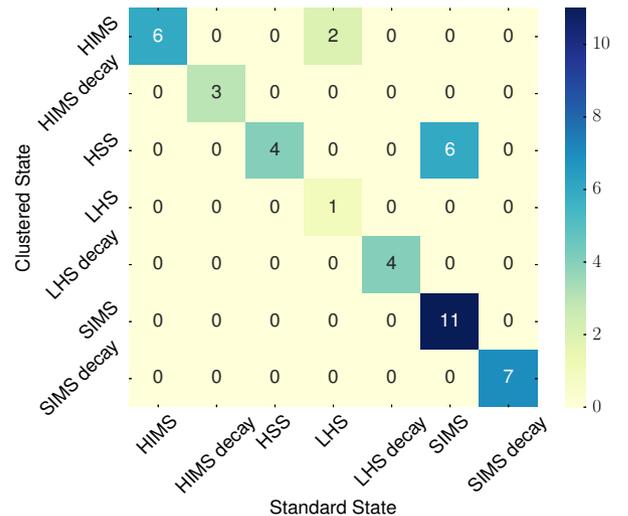}
	\end{center}
	\caption{The standard results are shown along the X-axis and K-Means clustering results are shown along the Y-axis of the confusion matrix. Diagonal entries denote correct or matching classification whereas off-diagonal entries denote a mismatch
		in classification. The ratio of diagonal counts to the total counts is the percentage match between the two classifications. The colour bar varies from yellow to blue as the counts increases.}
	\label{fig:heatmap}
\end{figure}

We start with the comparison of accretion state clustering using K-Means with the `standard' classification carried out for the same outburst (1999) of the source XTE J1859+226. In order to determine the number of clusters required, we make use of the WCSS plot (the elbow method, see Figure \ref{fig:k-Means_WCSS}). On visual inspection of this diagram, the elbow corresponds to an X-value (Number of clusters) in between 6 to 9. The bottom panel of Figure \ref{fig:k-Means_WCSS} shows the derivative of WCSS which flattens considerably as the number of clusters approaches 8, indicating that the rate of change of WCSS reduces significantly around 8 clusters. It suggests an optimum number of clusters for the classification to be equal to 8. However, as we are aware of the seven canonical states of BH-XRBs we choose the number of clusters (K) to be seven.
The normalised parameters along with the calculated EVRs are MJD (0.139), Total flux (0.096), Hardness Ratio (0.099), Photon index (0.060), Inner Disk Temperature (0.106) and presence of Type-C QPOs (0.248). The calculated EVRs for this outburst are mentioned in Table \ref{tab:EVR}.


\begin{figure*}
	\begin{center}
		\includegraphics[trim=0 0 0.25mm 0, clip = true, width=0.45\textwidth, height=6.5cm]{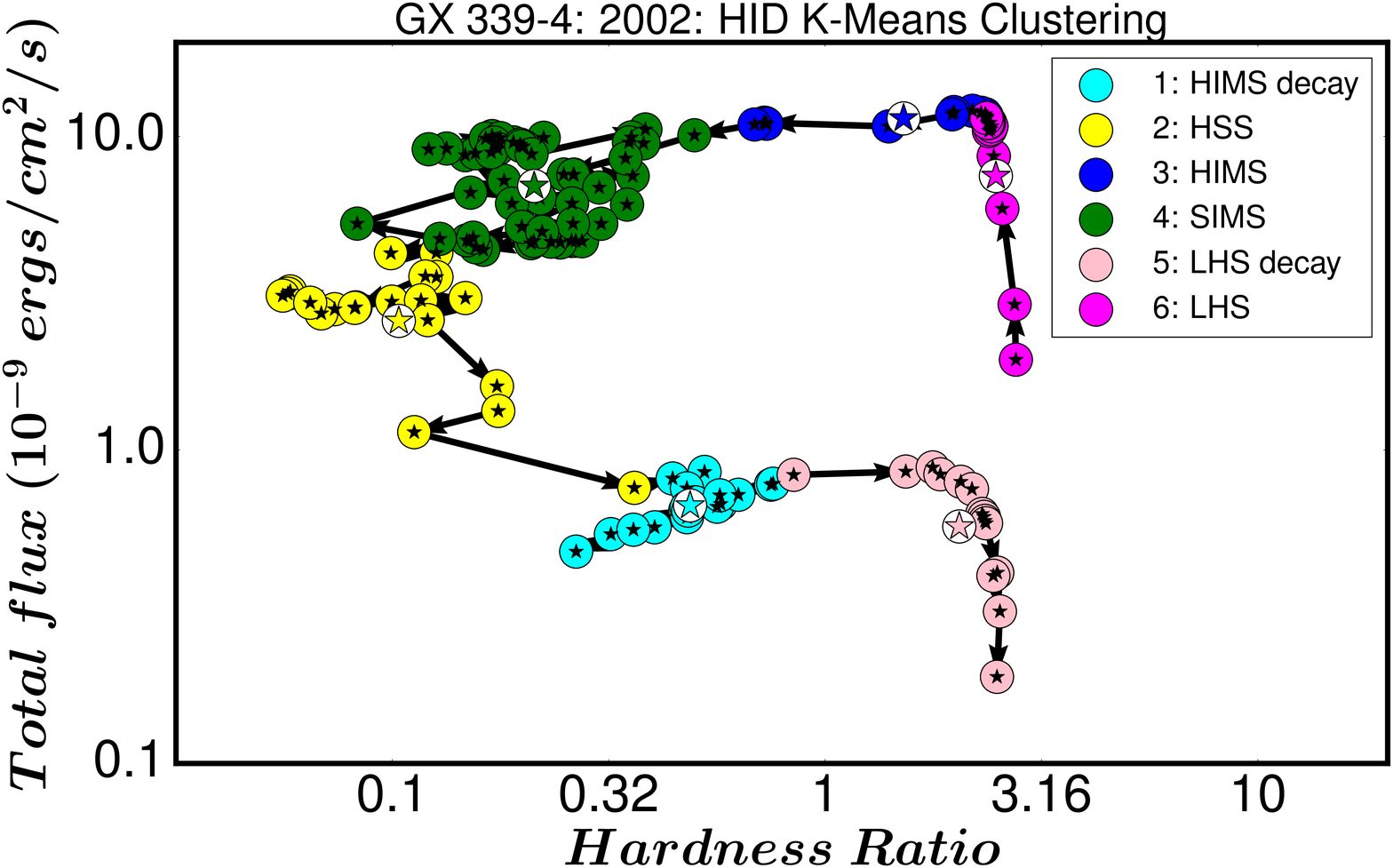}\
		\includegraphics[trim=0 0 0.25mm 0, clip = true, width=0.45\textwidth, height=6.5cm]{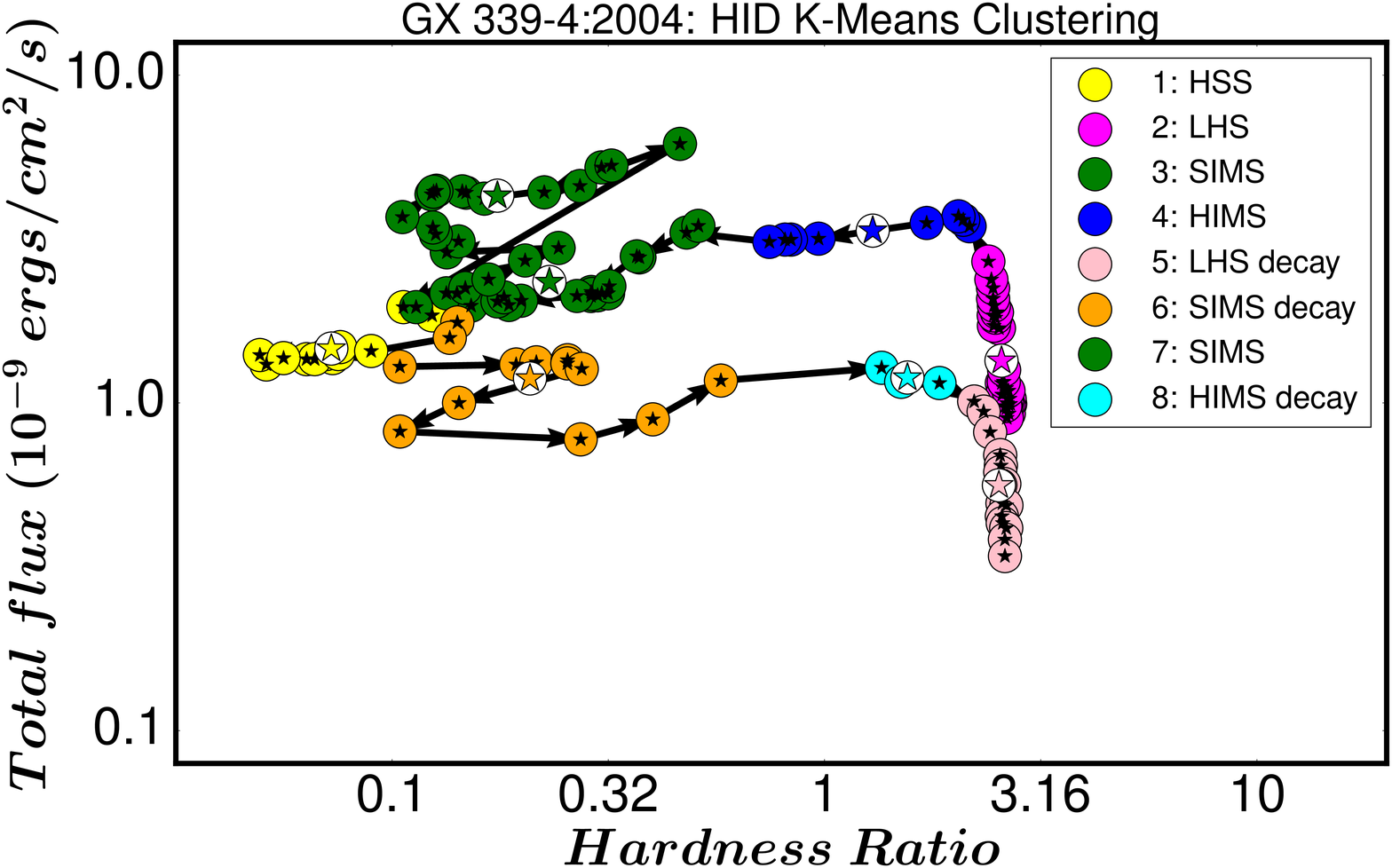}\\
		\includegraphics[trim=0 0 0.25mm 0, clip = true, width=0.45\textwidth, height=6.5cm]{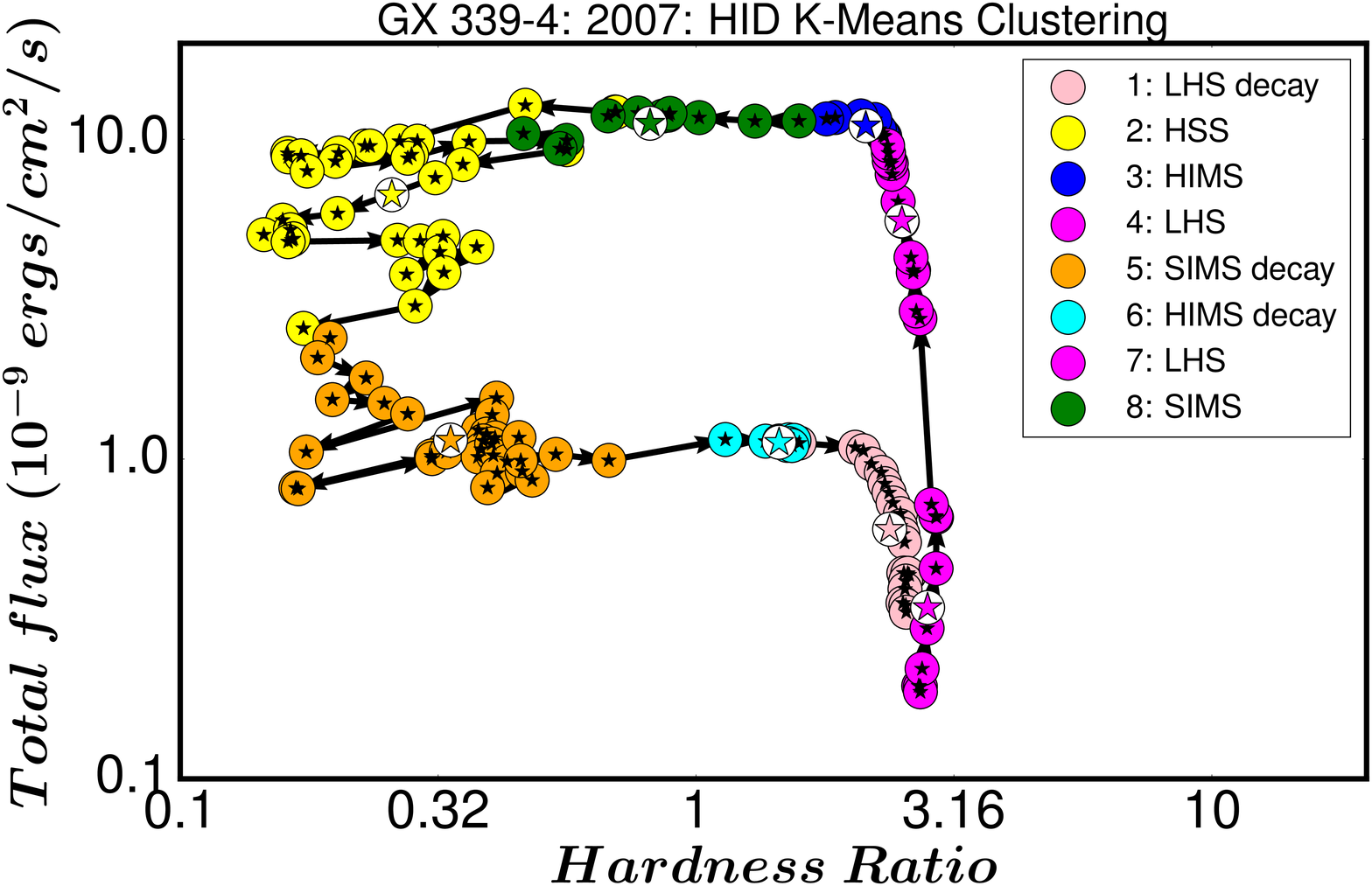}\
		\includegraphics[trim=0 0 0.25mm 0, clip = true, width=0.45\textwidth, height=6.5cm]{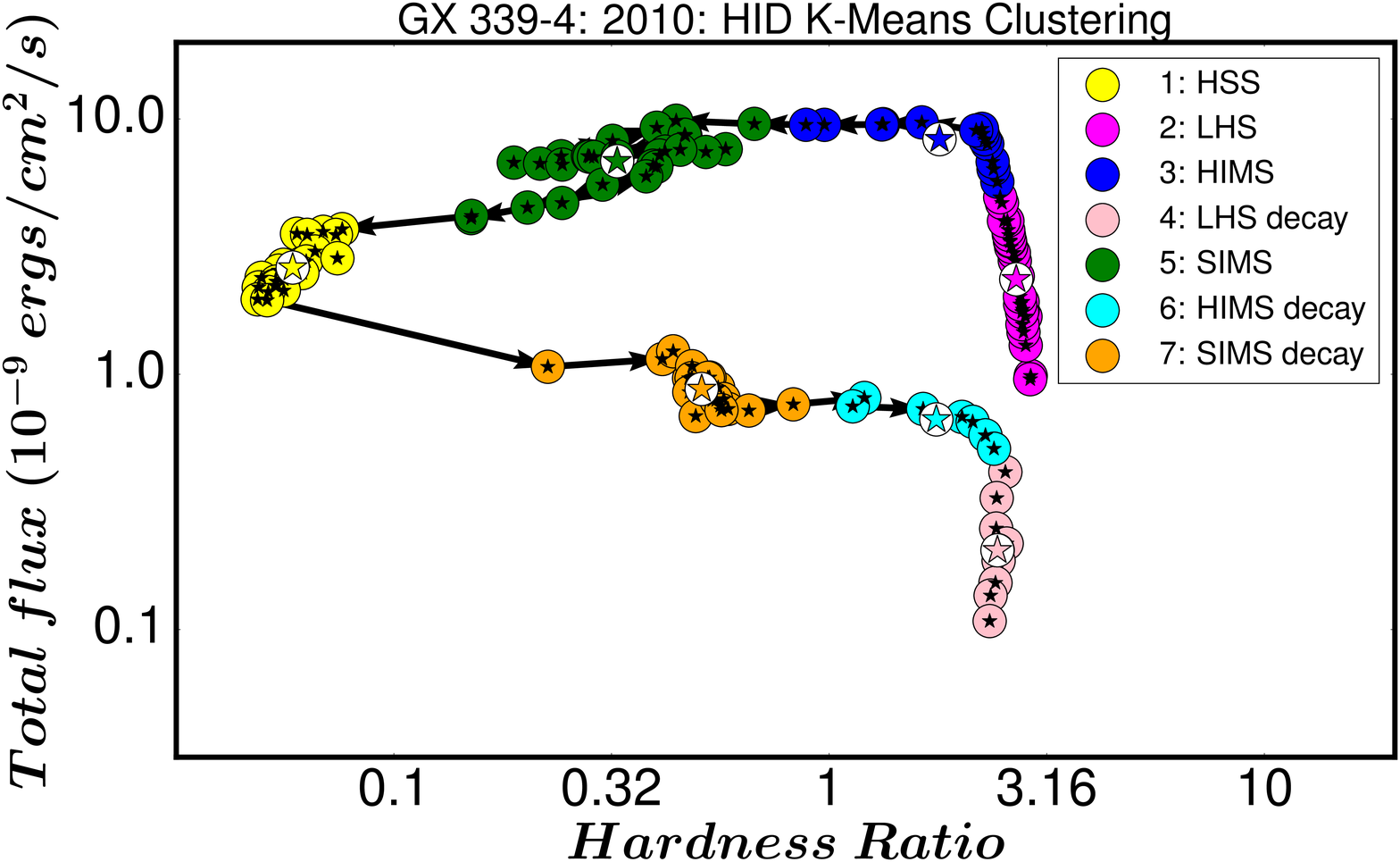}
	\end{center}
	\caption{Results of K-Means clustering of the q-diagram for multiple outbursts of GX 339-4 are presented with sequence as Top-left: 2002 outburst, Top-right: 2004 outburst, Bottom-left: 2007 outburst and Bottom-right: 2010 outburst.  Colour convention is same as that used in Figure \ref{fig:k-Means_XTE}. See text for details.}
	\label{fig:k-Means_GX}
\end{figure*}

\begin{figure*}
	\begin{center}
		
		\includegraphics[trim=0 0 0.25mm 0, clip = true, width=0.45\textwidth, height=6.5cm]{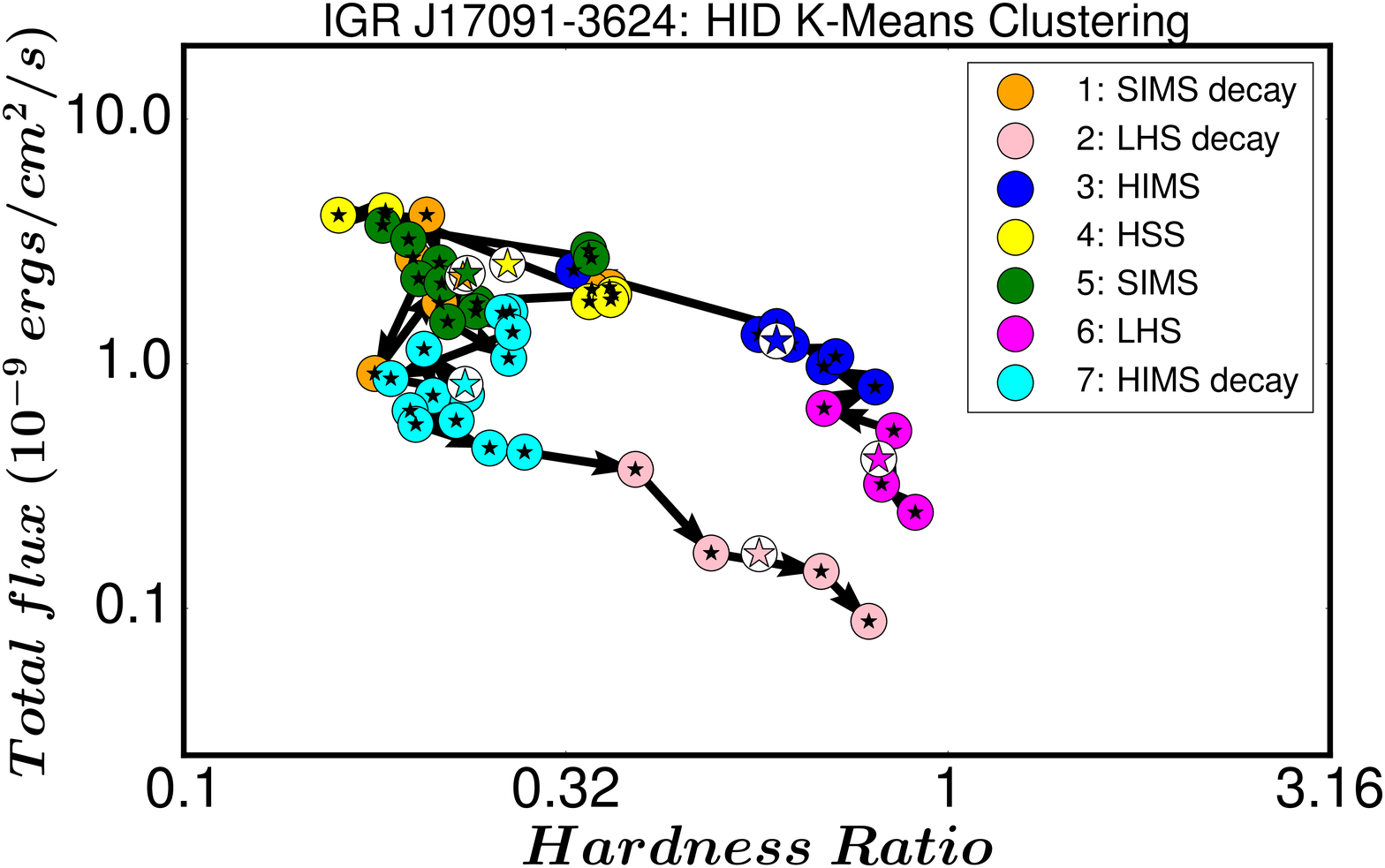}\
		\includegraphics[trim=0 0 0.25mm 0, clip = true, width=0.45\textwidth, height=6.5cm]{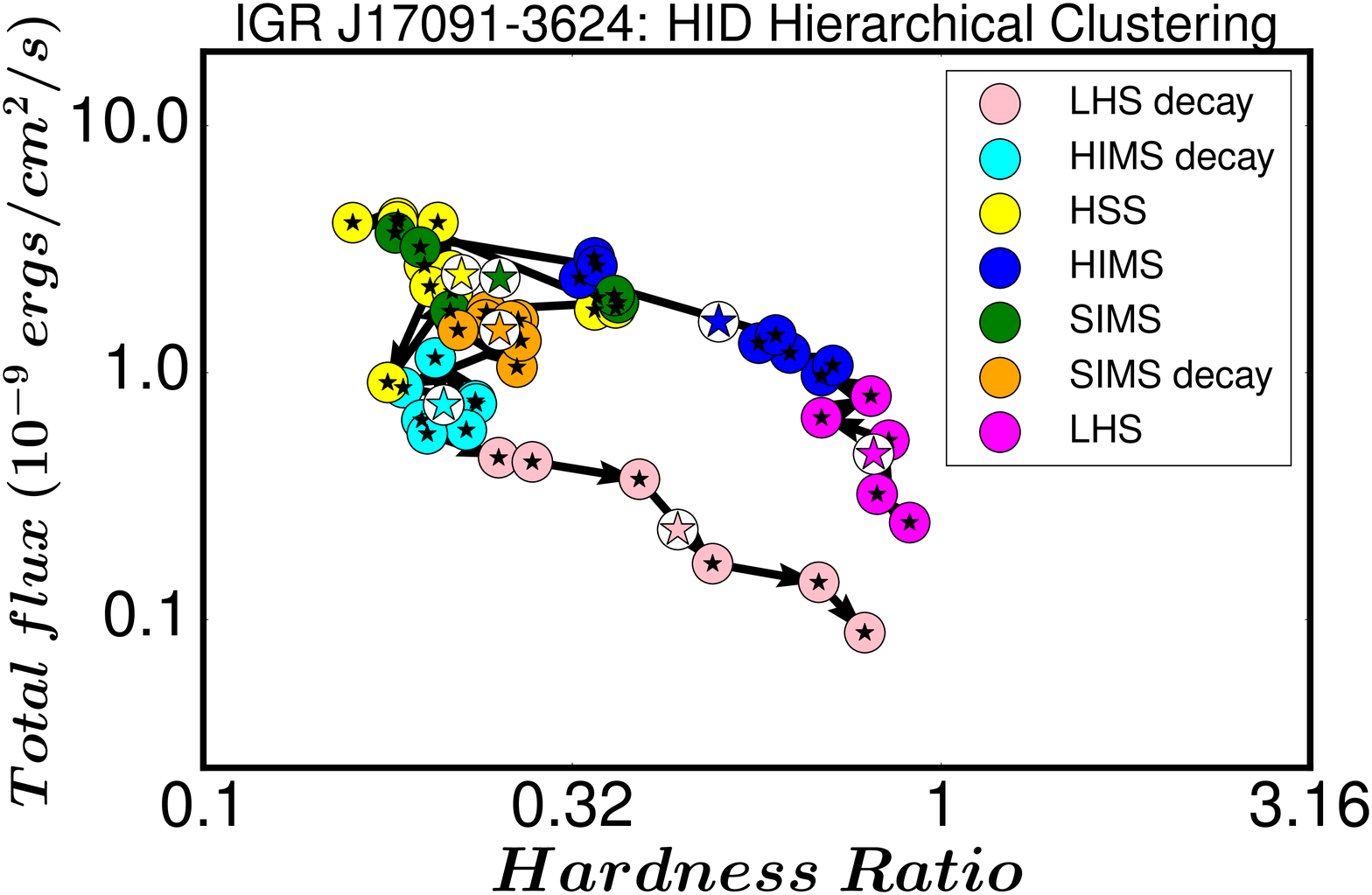}
		
	\end{center}
	\caption{Results of K-Means clustering of the q-diagram for the 2016 outburst of IGR J17091-3624 on the left and results of Hierarchical clustering of the same outburst on the right. Among the seven outbursts considered in this paper, this is the only case where Hierarchical clustering results have better accuracy than K-Means clustering when compared with the standard classification. See text for details.}
	\label{fig:k-Means_IGR}
\end{figure*}

Figure \ref{fig:k-Means_XTE} shows the output of K-Means clustering with 7 clusters. 
It indicates that the LHS of rising phase is an isolated observation unlike the standard method (left panel of Figure \ref{fig:XTEJparams}). The HIMS of the rising phase matches with the standard classification except for the initial two points.  The green coloured points in Figure \ref{fig:k-Means_XTE} specify the SIMS, which extends only for a small duration. This is because several points that were earlier classified as SIMS were more closely related to the properties of the yellow points indicating HSS. Following this, we have the decay phase with SIMS-decay (orange), HIMS-decay (cyan) and LHS-decay (pink). We bring to the reader's notice that the legend entries in each figure are presented in the order in which each cluster is detected. We use the same colour convention for marking each state in all the clustering results. The accuracy in classification is calculated as the ratio of diagonal entries to the total entries in the confusion matrix \citep{Fawcett2006}. The confusion matrix (see Figure \ref{fig:heatmap}) portrays the counts of predicted labels (Clustered States) versus the true labels (Standard States). It is evident that there are 8 off-diagonal entries and a total of 44 observations in the confusion matrix which corresponds to an accuracy of 82\%. This means 82\% of K-Means clustered states are correct when compared with the standard results. It must be noted that maximum discrepancy (6) between clustering results and standard results occurred with the classification of SIMS and HSS which generally happens to be the region where HR values fluctuate the most. Similar observations are also noted in the case of GX 339-4 and IGR J17091-3624 (see Figures \ref{fig:k-Means_GX} and \ref{fig:k-Means_IGR}). The precision (ratio of true positive detections to the total positive detections) for the K-Means classification is 91\% and the f1-score which is the harmonic mean of precision and sensitivity (ratio of true positive detections to actual positives) is 0.82. However, with hierarchical clustering only 45\% accuracy is obtained.

\subsection{GX 339-4: 2002, 2004, 2007 and 2010 Outbursts}
\label{s:Results_GX}

The BH-XRB source GX 339-4 was discovered by the Orbiting Solar Observatory-7 \citep{Markert1973}. 
During the {\it RXTE} era, it has gone into multiple outbursts in 1999, 2002, 2004, 2007 and 2010 \citep{Sreehari2019a}.
We use complete q-diagrams during 2002, 2004, 2007 and 2010 \citep{Belloni2005, Nandi2012, Aneesha2019} to classify the spectral states with the aid of K-Means clustering algorithm. The EVRs for the parameters considered for all outbursts are tabulated in Table \ref{tab:EVR}. The outcomes of K-Means clustering for this source are shown in Figure \ref{fig:k-Means_GX}. It is evident that GX 339-4 has completed the q-diagram in all the four outbursts considered.

Top-left panel of Figure \ref{fig:k-Means_GX} corresponds to the HID of 2002 outburst of GX 339-4. The clustering results have 90\% matching instances with the standard classification \citep{Belloni2005}. We have followed \cite{Belloni2005} by considering only six clusters for this outburst, as in the standard classification the source has not entered the SIMS decay state. 
The top-right panel corresponds to the 2004 outburst of GX 339-4. Though the source completes a q-profile, unlike other cases the flux of the source in HIMS is lower during this outburst. Moreover, it is the SIMS that has the peak flux during this outburst. The accuracy of classification using K-Means method for this outburst when compared to the standard results is 95\%. It should be noted that we have considered two regions as SIMS in this case. This was required to address the non-canonical behaviour of the source during the rising phase. 
Generally, we need to use only seven clusters. However, in the case of 2007 outburst, due to the data gap in the LHS of rising phase, the algorithm represents the LHS as two separate clusters. In order to overcome this problem we use an additional cluster which is also labelled as LHS. As standard classification details of this outburst are unavailable, we present the K-Means results without comparison. 
The q-diagram of the 2010 outburst has a canonical profile and is presented in the bottom-right panel of Figure \ref{fig:k-Means_GX}. The percentage of matching instances between K-Means clustering and standard classification \citep{Nandi2012} for this outburst is 87\%. Precision, accuracy and f1-score corresponding to these results are summarised in Table \ref{tab:match}. The accuracy score for the 2002, 2004, 2007 and 2010 outbursts of GX 339-4 with Hierarchical clustering are only 48\%, 47\%, 57\% and 54\% respectively.


\subsection{IGR J17091-3624: 2016 Outburst}
\label{s:Results_IGR}

The BH-XRB source IGR J17091-3624 was discovered in 2003 by the {\it INTEGRAL} observatory \citep{Kuulkers2003}. 
It has gone into outburst multiple times after its discovery. The source completes a q-profile during the 2016 outburst \citep{Radhika2018}. After calculating the EVRs, we use the K-Means clustering
algorithm on the HID (q-diagram) of its 2016 outburst. Left panel of Figure \ref{fig:k-Means_IGR} shows the result of clustering using the K-Means algorithm. There is only 57\% accuracy in this classification.  
On the right panel of Figure \ref{fig:k-Means_IGR}, we present the Hierarchical clustering results for the same outburst. 
Hierarchical clustering outcome shows 71\% match (accuracy) with standard classification. This is the only instance where hierarchical clustering gave a quantitatively better outcome than the K-Means method in the scenario of classification of accretion states. 
The precision for Hierarchical clustering outcome is 0.92 and f1-score is 0.77 for this outburst.

\subsection{MAXI J1535-571: 2017 Outburst}

\begin{figure}
	\begin{center}
		\includegraphics[trim=0 0 0.25mm 0, clip = true, width=0.45\textwidth, height=6.5cm]{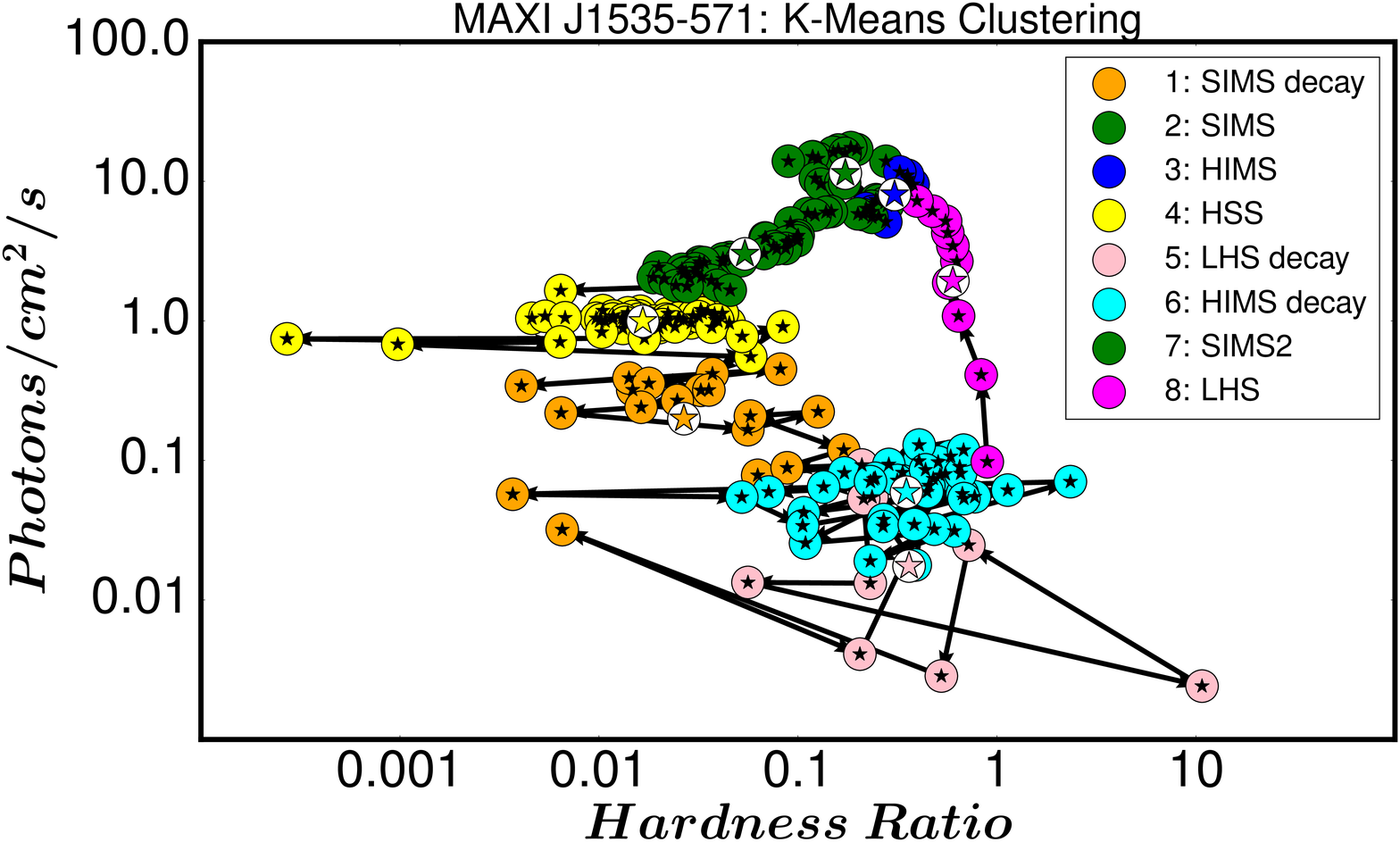}
	\end{center}
	\caption{Results of K-Means clustering of the q-diagram for the 2017 outburst of MAXI J1535-571. See text and Table \ref{tab:match} for details.}
	\label{fig:k-Means_MAXI}
\end{figure}

MAXI J1535-571 was discovered in 2017 simultaneously by {\it MAXI} \citep{Negoro2017} and {\it Swift} \citep{Kennea2017}
observatories. We performed both K-Means clustering and hierarchical clustering on the HID of this source generated using {\it MAXI} data. Figure \ref{fig:k-Means_MAXI} shows the different spectral states and the legend shows the corresponding colours. It should be noted that in the rising phase, the source goes through LHS -> HIMS --> SIMS --> HIMS2 --> SIMS2 --> HSS as is reported in \cite{Tao2018} and \cite{Sreehari2019}. In the decay phase, the source transits through SIMS-decay, HIMS-decay and finally to LHS-decay. The splits in HIMS were automatically detected by the algorithm. This is because, we have used QPO information from {\it AstroSat/LAXPC} and {\it Swift/XRT} as well  \citep{Sreehari2019}. The advantage here is that the decay phase data could be easily classified by clustering as opposed to the standard method wherein the large error bars on decay phase data made it difficult to classify the states. Clustering outcomes indicate 81\% match with standard results \citep{Sreehari2019} for the case of K-Means algorithm, while Hierarchical clustering results show only 58\% accuracy. The K-Means method has a precision of 89\% and f1-score equal to 0.82 when applied to this outburst, as summarised in Table \ref{tab:match}. It is to be noted that although unlike the other cases the HID is plotted in units of Photons/cm$^2$/s, the available data from {\it MAXI} and QPO information from {\it AstroSat} and {\it XRT} are enough to classify the states. So we have not looked into the archival data from {\it NICER} \citep{Miller2018} for this source.

\section{Discussion and Conclusion}
\label{s:Disc}
In this paper, we have applied Machine Learning based clustering algorithms in the scenario of accretion state classification of BH-XRBs for the first time. We have considered 7 outbursts from four different BH-XRB sources (XTE J1859+226, GX 339-4, IGR J17091-3624 and MAXI J1535-571) in order to check the feasibility of classifying accretion states using clustering algorithms. For this, we have considered two algorithms namely
K-Means clustering and Hierarchical clustering. The algorithms require the number of clusters as an input parameter.
The elbow method (see Figure \ref{fig:k-Means_WCSS}) indicates 8 clusters to be a reasonable choice. 
However, the optimal number of clusters required is known to be seven from standard classifications \citep{Belloni2005,Nandi2012}.
For some cases, we have used eight clusters as was required due to the presence of data gaps in observations (see bottom-left panel of Figure \ref{fig:k-Means_GX}) or due to non-canonical evolution of accretion state (see Figure \ref{fig:k-Means_MAXI}). Further, we pre-process the data by normalising all numerical parameters and applying `one-hot' encoding to categorical or non-numerical parameters like type of QPOs and instrument used for observation. 

The application of K-Means clustering for classification of spectral states for several outbursts as shown in this paper implies that it is a dependable method, especially when there are multiple criteria to be considered. The method has also proved its robustness
when applied to the {\it MAXI} data which has large uncertainties for the case of the 2017 outburst of the source MAXI J1535-571. The only disadvantage seems to be its inability to match clusters with similar characteristics when they are separated by data gaps as in the case of the 2007 outburst of GX 339-4 (see \S \ref{s:Results_GX}). This is because, the proximity in time (MJD) is one major factor for classification. When we try excluding the time (MJD) as a parameter, the algorithm classifies both LHS-rise (LHS) and LHS-decay into the same cluster. As we have to distinguish between these two states, we consider MJD as a necessary parameter.
The inclusion of categorical parameters like presence and type of QPOs and the instrument used which in turn specifies the energy range helps in carrying out a better classification. However, we have not made use of Type-B QPO information except for MAXI J1535-571. This is because though Type-B QPOs are usually observed in SIMS, they are not present in all SIMS observations \citep{Casella2005}. It implies that the presence of Type-B QPOs are only sufficient and not a necessary parameter for state classification.

We have also applied Hierarchical clustering for all seven outburst data. The percentage of matching classifications from Hierarchical clustering with `standard' results are lower than that obtained from K-Means clustering. Also, from visual inspection besides low precision and accuracy, it was clear that Hierarchical clustering results are sub-optimal in majority of the cases and hence we have not included those results in this paper. Reasonable outcomes with Hierarchical clustering are obtained only for 2016 outburst of IGR J17091-3624 (see \S \ref{s:Results_IGR} and Figure \ref{fig:k-Means_IGR}). Besides this, unlike K-Means clustering, Hierarchical clustering cannot correct itself as it proceeds in an agglomerative way \citep{Chattopad2007}. In Table \ref{tab:match}, we provide only the results from the quantitatively better (i.e. more accurate) of the two methods for each outburst.

\begin{table}
	\caption{The weighted average precision, accuracy and f1-score for K-Means clustering based classification with respect to the standard classification are presented in this table. The cases where Hierarchical clustering is used are denoted with $\dagger$ symbol. We point out that these values are not a measure of accuracy in classification, as we claim that the K-Means clustering based method gives more reliable results as is discussed in \S \ref{s:Disc}.}
	\begin{tabular}{|l|c|c|c|c|}
		\hline 
		Source & Outburst & Precision & Accuracy & f1-score \\
		\hline
		XTE~J1859+226   & 1999 & 0.91  &    0.82  &   0.82  \\
		
		GX~339-4        & 2002 & 0.93  &    0.90  &   0.90  \\
		
		GX~339-4        & 2004 & 0.96  &    0.95  &   0.95  \\	
		
		GX~339-4        & 2010 & 0.90  &    0.87  &   0.87  \\
		
		IGR~J17091-3624$\dagger$ & 2016 & 0.92  &    0.71  &   0.77 \\
		
		MAXI~J1535-571  & 2017 & 0.89  &    0.81  &   0.82 \\
		\hline
	\end{tabular}

	\label{tab:match}
\end{table}

Results obtained from K-means clustering method are consistent with the standard results (see Table \ref{tab:match}) and so it can be considered as an alternative to the `standard' approach.
However, pure observables like flux and hardness ratios are not enough for a good classification. 
As mentioned before, the presence of QPOs and spectral parameters like $\Gamma$ and ${\rm kT}_{\rm in}$ are also to be considered.
For instance, Type-C QPOs are usually found only in the LHS and HIMS states. Similarly, $\Gamma$ has low values in the harder states, while it is higher in the soft states as mentioned in the \S \ref{s:intro}. 
The weightage of these parameters are evident from the EVRs mentioned in Table \ref{tab:EVR}. In most of the cases the Type-C QPOs are found to have higher weightage than the other parameters. Anyway the values indicate that all the tabulated parameters are significant. The minimum weightage is for photon index with a value of 4.5\% for the 2010 outburst of GX 339-4. An alternative approach is to carry out principal component analysis (PCA) to generate new parameters called principal components from the existing parameters in the order of decreasing EVRs. However, we have not employed PCA in this work.

From the confusion matrix in Figure \ref{fig:heatmap}, it is clear that most of the confusion ($6/44=13.6\%$) in classification occurs in the transition from SIMS to HSS. Additionally, one may try using the fraction of disc emission as a parameter in order to address this confusion. 
As we have considered multiple parameters including photon indices and disk temperature in addition to the hardness ratios, flux, MJD and the presence and type of QPOs, we argue that the results from K-Means clustering are more accurate than the `standard' method that has been carried out till date.
However, for the case of 2002 outburst of GX 339-4, we knew before hand that the source transited only through six spectral states \citep{Belloni2005}. We were able to obtain 90\% match between K-Means outcomes and standard classification using this information.  

Thus, we have applied two clustering algorithms to classify spectral states of black hole binaries.
Though the percentage of matching results between the K-Means clustering and the `standard' classification is in the range of 81\% to 95\% (see Table \ref{tab:match}), we do not consider this measure as an accuracy of the new classification. Rather, we argue that the results from the K-Means clustering algorithm are better as it clubs together the observations with the most similar characteristics in the parameter space. If we are following the `standard' approach of classification, it becomes exceedingly difficult as the number of parameters increase. This is because, it is manually impossible to consider more than three dimensions at a time and hence when the number of parameters increase we cannot make a proper classification without the aid of a machine. As a result the clustering methods presented in this paper have more utility when compared to the standard techniques. K-Means clustering algorithm gives dependable classification as long as the data is continuous. Large data gaps may result in wrong classification which may indicate the need to include more clusters than required. This can be addressed with continuous monitoring of the black hole sources once they enter into an outburst phase or a small  intervention like changing the number of clusters while addressing the clustering results. Further improvements in clustering methods like Kernel K-Means and Spectral clustering \citep[and references therein]{Andrew2001,Dhillon2004} can be explored and will be presented elsewhere.

We have thus demonstrated the applicability of clustering algorithms in X-ray astronomy to perform the classification of accretion states of BH-XRBs for the first time. Prior to this, \cite{Huppenkothen2017} applied principal component analysis followed by logistic regression in order to classify the variability classes of the source GRS 1915+105 into chaotic and stochastic states.
Additionally, clustering algorithms can be employed in other areas also, like the study of properties of different branches of Z and Atoll sources \citep[and references therein]{Hasinger1989,Agrawal2018,Agrawal2020}.

Finally, we conclude and summarise our findings from this work as,
\begin{itemize}
	\item K-Means clustering when compared to standard results gives better accuracy than Hierarchical clustering.
	\item Most of the discrepancy between standard results and K-Means outcomes occur during the SIMS and HSS states.   
	\item K-Means clustering algorithm helps us to categorise data more accurately than the standard method, because of its ability to club together data with similar properties in a multi-dimensional parameter space.	
\end{itemize}

\section*{Acknowledgments}
The authors thank the reviewer for the valuable suggestions that have improved the quality of the manuscript.
AN thank GH, SAG; DD, PDMSA and Director, URSC for encouragement and continuous support to carry out this research.
Authors thank Ravishankar B.T. for careful reading and comments on the manuscript. This research made use of the data obtained from HEASARC (NASA) and AstroSat archive (Astrobrowse) of Indian Space Science Data Center (ISSDC). 

\section*{Data Availability}
Data used for this work is obtained from the HEASARC ({\small \url{https://heasarc.gsfc.nasa.gov/cgi-bin/W3Browse/w3browse.pl}}) archive of NASA and the AstroSat archive ({\small \url{https://astrobrowse.issdc.gov.in/astro\_archive/archive}}) of the Indian Space Science Data Center (ISSDC).


\bibliography{refs_ML}

 \label{lastpage}

\end{document}